
\magnification \magstep1
\raggedbottom
\openup 4\jot
\voffset6truemm
\headline={\ifnum\pageno=1\hfill\else
\hfill {\it Euclidean Maxwell theory in the
presence of boundaries. II} \hfill \fi}
\centerline {\bf EUCLIDEAN MAXWELL THEORY IN THE}
\centerline {\bf PRESENCE OF BOUNDARIES. II}
\vskip 1cm
\centerline {\bf Giampiero Esposito$^{1,2}$,
Alexander Yu Kamenshchik$^{3}$,}
\centerline {\bf Igor V Mishakov$^{3}$ and
Giuseppe Pollifrone$^{4}$}
\vskip 1cm
\centerline {\it ${ }^{1}$Istituto Nazionale di Fisica Nucleare}
\centerline {\it Mostra d'Oltremare Padiglione 20,
80125 Napoli, Italy;}
\centerline {\it ${ }^{2}$Dipartimento di Scienze Fisiche}
\centerline {\it Mostra d'Oltremare Padiglione 19,
80125 Napoli, Italy;}
\centerline {\it ${ }^{3}$Nuclear Safety Institute}
\centerline {\it Russian Academy of Sciences}
\centerline {\it 52 Bolshaya Tulskaya, Moscow 113191, Russia;}
\centerline {\it ${ }^{4}$Dipartimento di Fisica, Universit\`a
di Roma ``La Sapienza"}
\centerline {\it and INFN, Sezione di Roma}
\centerline {\it Piazzale Aldo Moro 2, 00185 Roma, Italy.}
\vskip 4cm
\leftline {PACS numbers: 0370, 0460, 9880}
\vskip 100cm
\noindent
{\bf Abstract.} $\zeta$-function regularization is applied to
complete a recent analysis of the quantized
electromagnetic field in the presence of boundaries. The
quantum theory is studied by setting to zero on the boundary the
magnetic field, the gauge-averaging functional and hence the
Faddeev-Popov ghost field. Electric boundary conditions are
also studied. On considering two gauge functionals which involve
covariant derivatives of the 4-vector potential,
a series of detailed calculations
shows that, in the case of flat Euclidean 4-space bounded by
two concentric 3-spheres, one-loop quantum amplitudes are
gauge independent and their mode-by-mode evaluation agrees
with the covariant formulae for such amplitudes and coincides
for magnetic or electric boundary conditions. By contrast,
if a single 3-sphere boundary is studied,
one finds some inconsistencies,
i.e. gauge dependence of the amplitudes.
\vskip 100cm
\leftline {\bf 1. Introduction}
\vskip 0.3cm
\noindent
Despite the lack of a mathematically consistent theory of
quantum gravity, the elliptic boundary-value problems
occurring in quantum cosmology have recently shed new light
on the whole quantization programme for gauge fields and gravitation
in the presence of boundaries [1-4].
Boundary effects play a key role in the path-integral approach to
quantum gravity [1], in the problem of boundary conditions for the
quantum state of the universe [5], and in comparing different
quantization and regularization techniques for gauge fields and
gravitation [1-2]. The latter problem provides the main motivation
for our paper. In fact many efforts have been produced in the
literature to understand the relation between canonical and
manifestly gauge-invariant approaches to quantum field theories,
as well as to compare mode-by-mode and covariant formulae for
the evaluation of quantum amplitudes.

In particular, in some papers dealing with
the calculation of the scaling
factor of the wave function of the universe in quantum cosmology,
discrepancies were found between results obtained by covariant and
non-covariant methods. It is well-known that this scaling factor
coincides with the Schwinger-De Witt coefficient $A_{2}$ in the
heat-kernel expansion [6].
Moreover, this factor can be
calculated by using the generalized Riemann
$\zeta$-function technique [1].
Within this framework, the prefactor is expressed through
the $\zeta(0)$ value, while $\zeta'(0)$ yields the full expression
for the one-loop effective action.
It was noticed that, for fields with non-zero spin,
calculations of the Schwinger-De Witt coefficient $A_{2}$
[6] by using
general covariant formulae for Riemannian 4-manifolds with boundaries
[7] give results which differ from those obtained by using
$\zeta$-function technique when one restricts the theory
to its physical degrees of freedom [1].
An analogous phenomenon was noticed in [8], where
$\zeta(0)$ was calculated for gravitons on the full Riemannian
de Sitter sphere.
In [9] the hypothesis was put forward that the reason of
discrepancies lies in the impossibility to perform a 3+1
decomposition on the Riemannian 4-manifolds under consideration.
Indeed, this
decomposition is necessary for the separation of physical degrees of
freedom. Thus, the direct calculation of $\zeta(0)$ in terms of
physical degrees of freedom
seems to be inconsistent. The $\zeta(0)$ calculation
for fermionic fields was then carried out in [9] on the part of flat
Euclidean 4-space bounded by two concentric 3-spheres. It was shown that
the discrepancy disappears in this case.

However, to understand discrepancies for the
electromagnetic field and other gauge theories,
it is necessary to take
into account ghost modes and non-physical degrees of freedom,
whose contributions may
survive in a non-trivial background even in the unitary gauges.
We here
restrict ourselves to a mode-by-mode analysis of the scaling factor
in relativistic gauges
about flat Euclidean 4-space with one or two 3-sphere boundaries.
One then faces the following
problems, here described in the case of Euclidean Maxwell theory,
which is the object of our investigation.

(i) {\it Choice of background 4-geometry}. This can be flat
Euclidean 4-space, or a curved Riemannian manifold providing
the index of the Dirac operator vanishes and no further
obstructions to having a unique, smooth solution of the
classical, elliptic boundary-value problem can be found.
[Knowledge of the index of the Dirac operator ensures one
understands what happens for second-order elliptic operators
as well]

(ii) {\it Choice of boundary 3-geometry}. Motivated by quantum
cosmology, this is taken to be a 3-sphere, or two concentric
3-spheres. These choices are necessary to have a unique smooth
solution of the corresponding classical boundary-value problem
for fields of various spins, and to avoid singularities at the
origin of the background 4-geometry (see below).

(iii) {\it Choice of boundary conditions}. They can be magnetic,
which implies setting to zero on the boundary the magnetic field,
the gauge-averaging functional and hence the Faddeev-Popov ghost
field. They can also be electric, hence setting to zero on
the boundary the electric field, and leading to Neumann conditions
on the ghost [1].

(iv) {\it Choice of gauge-averaging functional}. Here we focus
on the gauge-averaging functional first proposed in [1-2]:
$\Phi_{E}(A) \equiv { }^{(4)}\nabla^{\mu}A_{\mu}-A_{0} \; {\rm Tr}
\; K$ ($K$ being the extrinsic-curvature tensor of the boundary),
and on the Lorentz functional $\Phi_{L}(A) \equiv
{ }^{(4)}\nabla^{\mu}A_{\mu}$, as a first
check of gauge independence of
quantum amplitudes in a mode-by-mode analysis of the quantized
electromagnetic field.
\vskip 0.3cm
\noindent
The mode-by-mode analysis is performed by relying on the familiar
expansions of the components of the 4-vector potential on a
family of 3-spheres centred on the origin [1-3], i.e.
$$
A_{0}(x,\tau)=\sum_{n=1}^{\infty}R_{n}(\tau)Q^{(n)}(x)
\eqno (1.1)
$$
$$
A_{k}(x,\tau)=\sum_{n=2}^{\infty}\Bigr[f_{n}(\tau)S_{k}^{(n)}(x)
+g_{n}(\tau)P_{k}^{(n)}(x)\Bigr]
\; \; \; \; {\rm for} \; {\rm all} \; k=1,2,3
\eqno (1.2)
$$
where $Q^{(n)}(x),S_{k}^{(n)}(x),P_{k}^{(n)}(x)$ are scalar,
transverse and longitudinal vector harmonics on $S^{3}$
respectively. Note that, however, normal and tangential
components of $A_{\mu}$ are only well-defined at the 3-sphere
boundary where $\tau=\tau_{+}$, since a unit normal vector
field inside matching the normal to $S^{3}$ at the boundary is
ill-defined at the origin of the coordinate system for flat
Euclidean 4-space [9]. Hence the geometrical meaning of
(1.1)-(1.2) as normal and tangential components of the
4-vector potential inside the 3-sphere boundary remains unclear,
unless one studies an elliptic boundary-value problem where
flat Euclidean 4-space is bounded by two concentric 3-spheres
of radii $\tau_{+}$ and $\tau_{-}$, say.
The results of our calculations show that a proper study of
non-physical degrees of freedom and ghosts (which do not compensate
each other), together with the consideration
of manifolds possessing a consistent 3+1 decomposition, eliminates the
discrepancies between covariant and non-covariant formalisms.

Our paper is organized as follows. Section 2 studies one-loop
amplitudes by choosing the gauge-averaging functional
$\Phi_{E}(A)$ (see above), and imposing magnetic
(or electric) boundary conditions on 3-spheres.
The one-boundary and
two-boundary problems are analyzed.
Section 3 repeats the investigation of section 2 in the case
of the gauge functional $\Phi_{L}(A)$.
Results and open problems are presented in section 4.
Details relevant for $\zeta(0)$ calculations are described
in the appendix.
\vskip 0.3cm
\noindent
{\bf 2. One-boundary and two-boundary problems
in the Esposito gauge}
\vskip 0.3cm
\noindent
In this section we first
evaluate $\zeta(0)$ for the electromagnetic
field on the flat 4-dimensional Euclidean background bounded by a
3-sphere. We choose the magnetic boundary conditions
described in the introduction and carry out our calculations in
the Esposito gauge [1-2]. For this purpose, we begin by studying
the coupled eigenvalue
equations for normal and longitudinal components of the
electromagnetic field obtained in [1-2]. They have the form
(hereafter we set to 1 the parameter appearing in the
Faddeev-Popov action (2.3) of [2])
$$
{\widehat A}_{n}\;g_{n}(\tau)
+ {\widehat B}_{n}\;R_{n}(\tau) = 0
\eqno (2.1a)
$$
$$
{\widehat C}_{n}\;g_{n}(\tau)
+ {\widehat D}_{n}\;R_{n}(\tau) = 0
\eqno (2.1b)
$$
where
$$
{\widehat A}_{n} \equiv
{d^{2} \over d\tau^{2}} + {1 \over \tau}\;{d \over d\tau}
-{(n^{2}-1) \over \tau^{2}} + \lambda_{n}
\eqno (2.2)
$$
$$
{\widehat B}_{n} \equiv -{(n^{2}-1) \over \tau}
\eqno (2.3)
$$
$$
{\widehat C}_{n} \equiv -{1 \over \tau^{3}}
\eqno (2.4)
$$
$$
{\widehat D}_{n} \equiv
{d^{2} \over d\tau^{2}} + {3 \over \tau}\;{d \over d\tau}
-{(n^{2}-1) \over \tau^{2}} + \lambda_{n}.
\eqno (2.5)
$$
To study non-physical modes it is convenient to diagonalize the
operator matrix
$$
\left(\matrix{{\widehat A}_{n}&{\widehat B}_{n}\cr
{\widehat C}_{n}&{\widehat D}_{n}\cr}\right).
$$
Hence we look for a diagonalized matrix in the form
$$
O_{ij}^{(n)} \equiv
\left(\matrix{1&V_{n}(\tau)\cr W_{n}(\tau)&1}\right)\times
\left(\matrix{{\widehat A}_{n}&{\widehat B}_{n}\cr
{\widehat C}_{n}&{\widehat D}_{n}\cr}\right)\times
\left(\matrix{1&\alpha_{n}(\tau)\cr
\beta_{n}(\tau)&1\cr}\right).
\eqno (2.6)
$$
The matrix $\left(\matrix{1&\alpha_{n}\cr \beta_{n}&1\cr}\right)$
creates the linear combinations of functions
$R_{n}(\tau)$ and $g_{n}(\tau)$
which can be found from decoupled equations, whilst the matrix
$\left(\matrix{1&V_{n}\cr W_{n}&1}\right)$
selects these decoupled equations.

Setting to zero the off-diagonal matrix elements of (2.6),
and defining $\nu \equiv + \sqrt{n^{2}-{3\over 4}}$,
one finds equations solved by $V_{n}=-\alpha_{n}$,
$W_{n}=-\beta_{n}$, where
$$
\alpha_{n}(\tau) = \left(-{1 \over 2} \pm \nu
\right)\;\tau
\eqno (2.7)
$$
$$
\beta_{n}(\tau) = {1 \over (\nu+1/2)(\nu-1/2)}
\;\left({1 \over 2} \pm
\nu \right)\;{1 \over \tau}.
\eqno (2.8)
$$
Choosing the pair of solutions with upper sign
for $\alpha_{n}$ and lower sign for $\beta_{n}$
[the opposite choice of signs gives the equivalent system
of operators whilst the choice of
coinciding signs for $\alpha_{n}$ and
$\beta_{n}$ implies the degenerate system of equations]
one finds general basis functions for $R_{n}(\tau)$ and
$g_{n}(\tau)$ in the form
$$
g_{n}(\tau) = C_{1}\;I_{\nu - 1/2}(\sqrt{\lambda} \tau)
+ C_{2}\; \Bigr(\nu - 1/2 \Bigr)\;
I_{\nu + 1/2}(\sqrt{\lambda} \tau)
\eqno (2.9)
$$
$$
R_{n}(\tau)={1 \over \tau}\; \left(C_{1}\;
{-1 \over (\nu+1/2)} \;
I_{\nu - 1/2}(\sqrt{\lambda} \tau)
+ C_{2}\;I_{\nu + 1/2}(\sqrt{\lambda} \tau)\right)
\eqno (2.10)
$$
where $C_{1}$ and $C_{2}$ are constants.

In our particular gauge,
magnetic boundary conditions imply
Dirichlet boundary conditions for $g_{n}(\tau)$ and Neumann boundary
conditions for $R_{n}(\tau)$ [1-2].
The resulting equations lead to a $2 \times 2$ matrix,
hereafter denoted by $\cal I$, whose
determinant has to vanish to find non-trivial solutions. Such an
eigenvalue condition for normal and longitudinal components of
the electromagnetic potential is best studied by using the
algorithm of [4]. It is known that $\zeta(0)$ can be expressed as
$$
\zeta(0) = I_{\rm log} + I_{\rm pole}(\infty) - I_{\rm pole}(0)
\eqno (2.11)
$$
where [4]
$$
I(M^{2}, s) \equiv
\sum_{n=n_{0}}^{\infty} d(n)\; n^{-2s}
\; \log f_{n}(M^{2})
={I_{\rm pole}(M^{2})\over s}+I^{R}(M^{2})+{\rm O}(s).
\eqno (2.12)
$$
With our notation,
$d(n)$ is the degeneracy of the eigenvalues
parametrized by the integer $n$,
and $f_{n}(M^{2})$ is the corresponding eigenvalue condition.
Moreover, on analytic continuation,
$I_{\log}=I_{\rm log}^{R}$ is the coefficient of $\log M$ from
$I(M^{2}, s)$ as $M \rightarrow \infty$, and
$I_{\rm pole}(M^{2})$ is the residue at $s=0$.
The condition ${\rm det} \; {\cal I}=0$
should be studied after eliminating fake roots
$M = 0$. To obtain that it is enough to divide
${\rm det} \; {\cal I}$ by the minimal
power of M occurring in the determinant. It is easy to
see by using the series expansion for modified Bessel functions
[10] that such a power is $M^{2\nu-1}$.

We begin with the calculation of $I_{\log}$ for normal and
longitudinal modes of the electromagnetic field together with ghosts.
Using uniform asymptotic expansions for modified Bessel functions
[10] we can see that the only terms in
the logarithm of ${\rm det} \; {\cal I}$ divided
by $M^{2\nu-1}$ which are proportional to $\log M$ have the form
$
-2\nu\;\log M
$
whilst the ghost eigenvalue condition, divided by $M^{\nu}$,
gives analogous terms
$
-(2\nu+1)\;\log M
$
which contribute to $I_{\rm log}$ with the opposite sign.
Hence we can write
$$
I_{\rm log} = \sum\limits_{n=2}^{\infty} {n^{2} \over 2}
=- {1 \over 2}
\eqno (2.13)
$$
where $n^{2}$ is the dimension of the irreducible representation for
scalar hyperspherical harmonics. Summation in (2.13) is carried
out by the method of $\zeta$-function
regularization (see for details [4]).
The infinite sum starts from
$n=2$ since effects of zero-modes
for ghosts and normal photons should be calculated separately
whilst the longitudinal $(n=1)$ photon is absent. In [1-2]
it was found that the contribution of the decoupled
normal mode with magnetic boundary conditions is
$$
\zeta(0)_{\rm decoupled\;mode} = -{1 \over 4}.
\eqno (2.14)
$$
It is easy to calculate the contribution
to $\zeta(0)$ resulting from ghost
zero-modes by substituting into the corresponding expression for
$I_{\rm log}$ the value $n=1$. One finds
$$
\zeta(0)_{\rm ghost\; zero-modes} = 1.
\eqno (2.15)
$$

Now we have to calculate $I_{\rm pole}(\infty)$ and
$I_{\rm pole}(0)$. As shown in the appendix,
the structure of the term generating $I_{\rm pole}(\infty)$ is
$
{2\;\nu \over (\nu + 1/2)}.
$
Taking the logarithm of this expression and expanding it in inverse
powers of $n$ we can pick out the coefficient of the term $1/n$
in the expression
$
{n^{2} \over 2}\;\log {2\;\nu \over (\nu + 1/2)}
$
as $n \rightarrow \infty$, and find
$$
I_{\rm pole}(\infty)
=-{11 \over 96}.
\eqno (2.16)
$$
Now we can calculate $I_{\rm pole}(0)$ simply by using the usual
series expansion of Bessel functions [10] in
the limit  $M \rightarrow 0$.
The calculation relies on
the Stirling formula for the $\Gamma$-function appearing in
the series expansion for Bessel functions [10].
Omitting the details we write the result for $I_{\rm
pole}(0)$ for normal and longitudinal
modes of the electromagnetic field
together with ghosts (see appendix)
$$
I_{\rm pole\;(nonphys\;and\;ghosts)}(0)
=-{5 \over 12}.
\eqno (2.17)
$$
Now combining the results (2.13)-(2.17), one has
$$
\zeta(0)_{\rm normal,\;longitudinal,\;ghosts} = {53 \over 96}.
\eqno (2.18)
$$
Remarkably, the total contribution of
non-physical degrees of freedom and ghosts does not vanish.
Adding to (2.18) the contribution $-{77\over 180}$
of physical modes obtained in [3] one finds
$$
\zeta(0) = {179 \over 1440}.
\eqno (2.19)
$$

We now evaluate $\zeta(0)$ for the electromagnetic
field on the flat 4-dimensional Euclidean background bounded by
two concentric 3-spheres.
To begin with, let us calculate the contribution
to $\zeta(0)$ of physical degrees of freedom.
The basis functions for them are now the linear combination
$
f_{n}(\tau) = C_{1}\;I_{n}(M\tau) + C_{2}\;K_{n}(M\tau)
$
which should vanish at the 3-sphere boundaries of radii $\tau_{+}$
and $\tau_{-}$ respectively, where $\tau_{+} > \tau_{-}$.
This leads to the eigenvalue condition
$$
I_{n}^{-}\; K_{n}^{+} - I_{n}^{+}\;K_{n}^{-} = 0
\eqno (2.20)
$$
where $I_{n}^{-} \equiv  I_{n}(M\tau_{-}), I_{n}^{+} \equiv
I_{n}(M\tau_{+}), K_{n}^{-} \equiv  K_{n}(M\tau_{-}),
K_{n}^{+} \equiv
K_{n}(M\tau_{+})$.
Using series expansions for modified Bessel functions one can see
that the eigenvalue condition (2.20) has no fake roots. Since
in (2.20) the coefficients of products of
Bessel functions are independent of $n$, one finds that $I_{\rm
pole}(\infty) = 0$.
Bearing in mind that the dimension of irreducible representations for
transverse vector hyperspherical harmonics
is $2\;(n^{2}-1)$ [3],
where $n=2, \dots$, and using the hypergeometric expansions
for $I_{n}$ and $K_{n}$, one can easily show that
$
I_{\rm pole}(0) = 0.
$
Using the asymptotic expansions for $I_{n}$ and $K_{n}$ one can also
calculate $I_{\log}$ as
$$
I_{\log} = -\sum\limits_{n=2}^{\infty}
(n^{2} - 1)
= -{1 \over 2}
\eqno (2.21)
$$
and correspondingly
$$
\zeta(0)_{\rm transversal\; photons}=-{1 \over 2}.
\eqno (2.22)
$$
It is not difficult to consider also the contribution of ghosts. In this
case one has an eigenvalue condition coinciding with (2.20)
if we use $\nu \equiv + \sqrt{n^{2} - 3/4}$ instead of $n$.
Hence all 3 contributions to $\zeta(0)$ vanish, which implies
$$
\zeta(0)_{\rm ghosts} = 0.
\eqno (2.23)
$$

The next problem is the calculation of the contribution
to $\zeta(0)$ of the decoupled
normal mode, which has the form [1-2]
$
R_{1}(\tau) = C_{1}\;{1 \over \tau}\;I_{1}(M\tau) +
C_{2}\;{1 \over \tau}\;K_{1}(M\tau).
$
The derivative of this function should
vanish at the 3-sphere boundaries.
The determinant of the corresponding $2 \times 2$ matrix
should vanish and this equality gives,
as in the previous cases, the eigenvalue condition.
Such a determinant has no fake roots. Thus, by using the
uniform asymptotic expansions of Bessel functions one can
see that the $I_{\rm log}$ value is $-{1\over 2}$. However,
since our decoupled mode is non-vanishing for
$\tau \in [\tau_{-},\tau_{+}]$, one deals with a zero
eigenvalue corresponding to a non-zero eigenfunction
satisfying boundary conditions.
The number $N_{D}=1$ of such decoupled normal modes contributes to
the full $\zeta(0)$ value. Hence one finds
$$
\zeta(0)_{\rm decoupled\;mode}
=I_{\rm log}+N_{D}=-{1\over 2}+1
={1 \over 2}.
\eqno (2.24)
$$

We now evaluate
the contribution of the coupled normal and longitudinal modes of
the electromagnetic field.
Since in the two-boundary case the singularity at the origin is
avoided, both $I$-
and $K$-functions contribute to gauge modes.
Hence the general solutions for $g_{n}(\tau)$
and $R_{n}(\tau)$ are
$$ \eqalignno{
g_{n}(\tau)&= C_{1}\;I_{\nu - 1/2}(M \tau)
+ C_{2}\;(\nu - 1/2)\;
I_{\nu + 1/2}(M \tau) \cr
&+ C_{3}\;K_{\nu - 1/2}(M \tau)
+ C_{4}\;(\nu - 1/2)\;
K_{\nu + 1/2}(M \tau)
&(2.25)\cr}
$$
$$ \eqalignno{
R_{n}(\tau)& ={1 \over \tau}\; \left(C_{1}\;
{-1 \over (\nu+1/2)}\;
I_{\nu - 1/2}(M \tau)
+ C_{2}\;I_{\nu + 1/2}(M \tau)\right. \cr
&+ \left. C_{3}\;
{-1 \over (\nu+1/2)}\;
K_{\nu - 1/2}(M \tau)
+ C_{4}\;K_{\nu + 1/2}(M \tau)\right).
&(2.26)\cr}
$$
After substitution of (2.25)-(2.26) into magnetic boundary
conditions at the 3-sphere boundaries one has a system of
4 equations. The determinant of the corresponding
$4 \times 4$ matrix should vanish. In such a determinant,
the smallest power of $M$ is $M^{-2}$.
Thus, to avoid the appearance of fake roots in
the eigenvalue condition, it is necessary to multiply it by
$M^{2}$. Taking into account the formulae for the asymptotic
expansions of Bessel functions one can see that the
coefficient of $\log M$ in the logarithm of our eigenvalue condition
vanishes, hence $I_{\log}$ vanishes as well. Let us now calculate
$I_{\rm pole}(\infty)$. Just as in the previous problem,
this value is determined by $n$-dependent coefficients in the
determinant and can be calculated from the expression
$
{n^{2}\over 2}\log
\left({4 \nu^{2} \over (\nu + 1/2)^{2}}\right)
$
along the lines described in the appendix. One has
$
I_{\rm pole}(\infty)
=-{3\over 16}-{1\over 24}
= -{11 \over 48}.
$

Now using the usual hypergeometric expansions for modified Bessel
functions one obtains
$
I_{\rm pole}(0) = - {11 \over 48},
$
and since the values of $I_{\rm pole}$
at $\infty$ and at $0$ compensate
each other, and $I_{\log} = 0$, one finds
$$
\zeta(0)_{\rm normal\;and\;longitudinal} = 0.
\eqno (2.27)
$$

Combining the results (2.22)-(2.24) and (2.27) one has
$$
\zeta(0) = 0
\eqno (2.28)
$$
which coincides with the covariant result.
Of course, the covariant $\zeta(0)$ value is zero, since the volume
contribution on the flat background vanishes whilst two-boundary
contributions compensate each other (look at the corresponding
formulae in [7]).

On imposing electric boundary conditions, which are motivated
by supersymmetric quantum field theory [1], the situation
for coupled normal and longitudinal modes is
just opposite to that in the magnetic case, since normal modes
and the normal derivatives of longitudinal modes vanish at the
3-sphere boundaries [1-2]. Defining $a_{\nu} \equiv
{-1\over (\nu+1/2)}$, $b_{\nu} \equiv (\nu-1/2)$,
the corresponding eigenvalue condition is the vanishing of the
determinant of the matrix
$$
\left(\matrix{
a_{\nu}I_{b_{\nu}}^{-}&
I_{b_{\nu}+1}^{-}&
a_{\nu}K_{b_{\nu}}^{-}&
K_{b_{\nu}+1}^{-}\cr
a_{\nu}I_{b_{\nu}}^{+}&
I_{b_{\nu}+1}^{+}&
a_{\nu}K_{b_{\nu}}^{+}&
K_{b_{\nu}+1}^{+}\cr
(I_{b_{\nu}-1}^{-}+I_{b_{\nu}+1}^{-})&
b_{\nu}(I_{b_{\nu}}^{-}+I_{b_{\nu}+2}^{-})&
-(K_{b_{\nu}-1}^{-}+K_{b_{\nu}+1}^{-})&
-b_{\nu}(K_{b_{\nu}}^{-}+K_{b_{\nu}+2}^{-})\cr
(I_{b_{\nu}-1}^{+}+I_{b_{\nu}+1}^{+})&
b_{\nu}(I_{b_{\nu}}^{+}+I_{b_{\nu}+2}^{+})&
-(K_{b_{\nu}-1}^{+}+K_{b_{\nu}+1}^{+})&
-b_{\nu}(K_{b_{\nu}}^{+}+K_{b_{\nu}+2}^{+})\cr}\right).
$$
This yields
$$
I_{\log} = 0 \; \; \; \; I_{\rm pole}(\infty) = I_{\rm pole}(0)
=-{11\over 48}
\eqno (2.29)
$$
which implies
$$
\zeta(0)_{\rm normal\;and\;longitudinal} = 0.
\eqno (2.30)
$$
Hence one finds
$$
\zeta(0) = 0.
\eqno (2.31)
$$
Such a $\zeta(0)$ value
coincides with the covariant result (as in the case of magnetic
boundary conditions, the contributions
of two 3-sphere boundaries in covariant
formalism cancel each other).

It is easy to carry out the corresponding calculations in the
Lorentz gauge (section 3). They yield again $\zeta(0)=0$.
Thus, we have shown that in the case of electric boundary conditions
(just as in the case of magnetic ones) on the
flat Riemannian 4-manifold with two 3-sphere
boundaries leading to a well-defined 3+1 split
of the 4-vector potential,
results of covariant and mode-by-mode formalisms coincide.
\vskip 0.3cm
\noindent
{\bf 3. One-boundary and two-boundary problems in the
Lorentz gauge}
\vskip 0.3cm
\noindent
In the Lorentz gauge, coupled eigenvalue equations for
normal and longitudinal modes are again described by a system of
the kind (2.1a)-(2.1b). However, the four operators are
replaced by other operators ${\widehat E}_{n},
{\widehat F}_{n},{\widehat G}_{n}, {\widehat H}_{n}$
respectively, where the operator ${\widehat E}_{n}$
coincides with ${\widehat A}_{n}$ written in section 2,
whilst ${\widehat F}_{n},\; {\widehat G}_{n}$
and ${\widehat H}_{n}$ are
different and take the form
$$
{\widehat F}_{n} \equiv {2\;(n^{2} - 1) \over \tau}
\eqno (3.1a)
$$
$$
{\widehat G}_{n} \equiv {2 \over \tau^{3}}
\eqno (3.1b)
$$
$$
{\widehat H}_{n} \equiv
{d^{2} \over d\tau^{2}} + {3 \over \tau}\;{d \over d\tau}
-{(n^{2} + 2) \over \tau^{2}} + \lambda_{n}.
\eqno (3.1c)
$$
The corresponding system of equations
can be diagonalized in the same way as the analogous
system (2.1a)-(2.1b). Hence one finds a
couple of Bessel-type equations, giving the general
solutions for $g_{n}(\tau)$ and $R_{n}(\tau)$
in the one-boundary problem as
$$
g_{n}(\tau) = C_{1}\;I_{n+1}(M\tau)+C_{2}\;(n+1)\;I_{n-1}(M\tau)
\eqno (3.2)
$$
$$
R_{n}(\tau) = {1\over \tau}\Bigr(
-C_{1}\;{1 \over (n - 1)}\;I_{n+1}(M\tau) +
C_{2}\;I_{n-1}(M\tau)\Bigr).
\eqno (3.3)
$$
In the magnetic case we are
studying, one sets to zero on the boundary the gauge-averaging
functional. If this is the Lorentz functional, one is choosing
Dirichlet conditions for $g_{n}$ modes and Robin conditions
for $R_{n}$ modes, i.e.
$
g_{n}(\tau_{+})=0,
{\dot R}_{n}(\tau_{+})+{3\over \tau_{+}}R_{n}(\tau_{+})=0
$ [1-2].
Such a system leads to the eigenvalue condition
$$
{\rm det} \left(\matrix{I_{n+1}^{+}&(n+1) I_{n-1}^{+}\cr
-{1 \over (n-1)}\left({2I_{n+1}^{+} \over (M\tau_{+}/2)}+
I_{n}^{+}+I_{n+2}^{+}\right)&
\left({2I_{n-1}^{+} \over
(M\tau_{+}/2)}+I_{n-2}^{+}+I_{n}^{+}\right)\cr}\right) = 0.
\eqno (3.4)
$$
To get rid of fake roots it is necessary to
divide the determinant (3.4) by $M^{2n-1}$, and then we can calculate
$I_{\log}$ as
$$
I_{\log} = -\sum\limits_{n=2}^{\infty} n^{3}
={119 \over 120}.
\eqno (3.5)
$$

$I_{\rm pole}(\infty)$ can be obtained by
extracting the $n$-dependent coefficients in the determinant (3.4),
which gives
$
{n^{2}\over 2} \log
\left({2\;n \over (n - 1)}\right).
$
{}From this expression one obtains (see appendix)
$$
I_{\rm pole}(\infty)
={1 \over 6}.
\eqno (3.6)
$$

Now we can calculate $I_{\rm pole}(0)$ by taking the logarithm of
the determinant (3.4) in the limit $M \rightarrow 0$ and expanding it
in inverse powers of $n$. The result is (see appendix)
$$
I_{\rm pole}(0)
={1\over 360}+{1\over 3}
={121 \over 360}.
\eqno (3.7)
$$
Combining together the results (3.5)-(3.7) one finds
$$
\zeta(0)_{\rm normal\;and\;longitudinal} = {37 \over 45}.
\eqno (3.8)
$$

The equation for the decoupled normal mode implies
$R_{1}={1 \over \tau}\;I_{2}(M\tau)$,
and the magnetic boundary condition leads to
$$
\zeta(0)_{\rm decoupled} = -{3 \over 4}.
\eqno (3.9)
$$

The eigenvalue condition for ghosts in the Lorentz
gauge coincides with the one for scalar
fields and the corresponding $\zeta(0)$ value can be
obtained from the well-known result
for a scalar field subject to Dirichlet boundary
conditions [1]. Taking into account the change of sign
due to the fermionic nature of ghosts corresponding to
spin-1 fields [1-2] one has
$$
\zeta(0)_{\rm ghosts} = {1 \over 90}.
\eqno (3.10)
$$
Combining the results (3.8)-(3.10) one obtains
$$
\zeta(0)_{\rm normal,\;longitudinal,\;ghosts} = {1 \over 12}.
\eqno (3.11)
$$
Correspondingly, the total
$\zeta(0)$ value including the contribution
$-{77\over 180}$ of physical modes is
$$
\zeta(0) = -{31 \over 90}.
\eqno (3.12)
$$
Remarkably, (3.12) agrees with the corrected $\zeta(0)$ value
obtained in [7], which relies on the analysis by Vassilevich
[11]. However, (3.12)
differs from the mode-by-mode result (2.19).
This discrepancy seem to originate
from the ill-definiteness of the 3+1 decomposition of the 4-vector
potential on the manifold under consideration [9].

In the two-boundary case,
the eigenvalue condition for the coupled normal and longitudinal
modes in the Lorentz gauge on the flat
Riemannian 4-manifold with two concentric 3-sphere boundaries is
(on imposing magnetic boundary conditions)
$$
\det \left(\matrix{I_{n+1}^{-}&(n+1) I_{n-1}^{-}&
K_{n+1}^{-}&(n+1) K_{n-1}^{-}\cr
I_{n+1}^{+}&(n+1) I_{n-1}^{+}&
K_{n+1}^{+}&(n+1) K_{n-1}^{+}\cr
-{1 \over (n-1)}\bigl({2I_{n+1}^{-} \over M\tau_{-}/2} \atop
+I_{n}^{-}+I_{n+2}^{-}\bigr)&
\bigl({2I_{n-1}^{-} \over M\tau_{-}/2}
\atop +I_{n-2}^{-}+I_{n}^{-}\bigr)&
-{1 \over (n-1)}\bigl({2K_{n+1}^{-} \over M\tau_{-}/2} \atop
-K_{n}^{-}-K_{n+2}^{-}\bigr)&
\bigl({2K_{n-1}^{-} \over M\tau_{-}/2}
\atop -K_{n-2}^{-}-K_{n}^{-}\bigr)\cr
-{1 \over (n-1)}\bigl({2I_{n+1}^{+} \over M\tau_{+}/2} \atop
+I_{n}^{+}+I_{n+2}^{+}\bigr)&
\bigl({2I_{n-1}^{+} \over M\tau_{+}/2}
\atop +I_{n-2}^{+}+I_{n}^{+}\bigr)&
-{1 \over (n-1)}\bigl({2K_{n+1}^{+} \over M\tau_{+}/2} \atop
-K_{n}^{+}-K_{n+2}^{+}\bigr)&
\bigl({2K_{n-1}^{+} \over M\tau_{+}/2}
\atop -K_{n-2}^{+}-K_{n}^{+}\bigr)\cr}
\right) = 0.
\eqno (3.13)
$$
The contribution to $I_{\log}$ of the determinant (3.13) is $0$.
The function determining the behaviour of $I_{\rm pole}(M^{2})$
as $M \rightarrow \infty$ and $n \rightarrow \infty$ has
the form
$
{n^{2}\over 2} \log
\left({4 n^{2} \over (n - 1)^{2}}\right)
$
and correspondingly
$
I_{\rm pole}(\infty) = {1 \over 3}.
$
After taking the logarithm of the determinant (3.13) at $M = 0$ and
expanding it in inverse powers of $n$ one obtains
$
I_{\rm pole}(0) = {1 \over 3}.
$
Thus, one finds again compensation of $I_{\rm pole}(\infty)$ and
$I_{\rm pole}(0)$ and
$
\zeta(0)_{\rm normal\;and\;longitudinal} = 0.
$
Finally one obtains the total result
$$
\zeta(0) = 0
\eqno (3.14)
$$
coinciding with that obtained in section 2 and with the covariant
one.
\vskip 0.3cm
\leftline {\bf 4. Concluding remarks}
\vskip 0.3cm
\noindent
The main results of our investigation are as follows.

First, we have proved that, in the case of flat Euclidean
4-space bounded by two concentric 3-spheres, the mode-by-mode
analysis of one-loop quantum amplitudes for Euclidean Maxwell
theory agrees with the covariant formulae used in [7].
Moreover, such Faddeev-Popov quantum amplitudes are indeed
gauge independent in the cases studied in our paper
(see below).
Second, we have shown that contributions of gauge modes and
ghost fields to the full $\zeta(0)$ value do not cancel
each other.
Hence the reduction of a gauge theory to its physical
degrees of freedom before quantization only yields the
contribution of such degrees of freedom to the quantum theory,
but is by itself insufficient to describe a gauge-invariant
quantum theory.
Third, we have provided evidence that, when the boundary
3-geometry does not lead to a well-defined 3+1 split of
the 4-vector potential, some inconsistencies occur, i.e.
one-loop quantum amplitudes are gauge dependent.

Interestingly, this seems to complement the Hartle-Hawking
programme [5], which relies on a Wick-rotated path integral with
just one boundary 3-geometry. More precisely,
on the one hand we know that the use of
unitary gauge, 3+1 decomposition, Hamiltonian formalism and
extraction of physical degrees of freedom is necessary to recover
the physical content of the theory. Moreover, the natural way of
implementing the Hartle-Hawking programme [5]
involves the consideration of the wave function
of the universe in the Lorentzian
region in terms of physical degrees of freedom
(with the subsequent analytic continuation
to the Euclidean-time region, whenever this is possible).
On the other hand,
the Euclidean germ from which our universe might originate [5]
is the Riemannian 4-manifold which does not
possess a well-defined 3+1 split [9].

Does this mean that it is impossible to carry out the Hartle-Hawking
programme in a consistent way ?
Indeed, it should be emphasized that our present
understanding of quantum field theory and quantum gravity does not
enable one to make a conclusive statement. Since
quantum amplitudes involve differential operators and their
eigenvalues, the mode-by-mode analysis based on $\zeta$-function
regularization remains of primary importance.
Moreover, the analytic continuation back to
real, Lorentzian time may be impossible [1]. Hence there are cases
where we may have to limit ourselves
to the elliptic boundary-value problems of
Riemannian geometry, where one cannot define the notion of
time-evolution.

Other interesting problems remain unsolved as well.
In fact, one still has to obtain a general proof of
gauge invariance in the two-boundary case. Our paper
has only focused on two {\it particular} gauge-averaging
functionals. Moreover, one has to repeat our mode-by-mode
analysis of one-loop amplitudes
for spin-${3\over 2}$ fields and gravitation,
as well as deal with curved background 4-geometries.
In the latter case, it is impossible to decouple gauge
modes without studying fourth-order ordinary differential
equations, and one faces the technical problem of working
out the uniform asymptotic expansions of their solutions.

The new tools developed in our paper
and in the recent literature [1-4,7,9,11]
make us feel that a complete mode-by-mode analysis of quantized
gauge fields and gravitation in the presence of boundaries
is in sight. Although this would be far from having a good
theory of quantum gravity, it seems to add evidence in favour
of quantum cosmology being at the very heart of modern quantum
field theory.
\vskip 0.3cm
\leftline {\bf Acknowledgments}
\vskip 0.3cm
\noindent
G Esposito is indebted to Peter D'Eath
and Jorma Louko for introducing him to the mode-by-mode
analysis of quantized Maxwell theory.
A Yu Kamenshchik and I V Mishakov are indebted to Andrei
Barvinsky for several enlightening conversations.
Anonymous referees made comments which led to a substantial
improvement of the original manuscript.
Our joint paper was supported in part by the European union
under the Human Capital and Mobility Program. Moreover, the
research described in this publication was made possible in
part by Grant No MAE000 from the International Science
Foundation. The work of A Kamenshchik was partially supported
by the Russian Foundation for Fundamental Researches through
grant No 94-02-03850-a.
\vskip 0.3cm
\leftline {\bf Appendix}
\vskip 0.3cm
\noindent
To clarify our $\zeta(0)$ calculations,
some of them are here presented
in more detail. We first evaluate the
contribution to $\zeta(0)$ of non-physical modes in the Lorentz gauge
for the manifold with one boundary (section 3). For this purpose,
let us write the determinant (3.4) giving the eigenvalue
condition in the explicit form
$$
I_{n+1}^{+} \left({2I_{n-1}^{+} \over (M \tau_{+}/2)} +
I_{n-2}^{+} + I_{n}^{+}\right)
+{(n+1) \over (n-1)}\;I_{n-1}^{+}
\left({2I_{n+1}^{+} \over (M \tau_{+}/2)} +
I_{n}^{+} + I_{n+2}^{+}\right)=0.
\eqno (A1)
$$
The minimal power of
$M$ in (A1) is $2n - 1$.
Thus, to get rid of the fake roots
$M=0$ we should divide our eigenvalue condition by $M^{2n-1}$.

The leading behaviour of $I_{n}(nM\tau)$ is
determined by the exponent $e^{nM\tau}$
multiplied by ${1 \over \sqrt{M\tau}}$.
Hence, after taking the logarithm
of (A1) divided by $M^{2n-1}$ one finds that the only term
proportional to $\log M$ is
$
(-2n) \log M.
$
Having this expression we can get, by direct summation, the result
(3.5) for $I_{\log}$.
For the calculation of $I_{\rm pole}(\infty)$ it is more convenient
to re-write (A.1) through derivatives of Bessel
functions instead of using the recurrence formulae. Thus we have,
instead of (A1), the equation
$$
I_{n+1}^{+} \left({I_{n-1}^{+} \over (M \tau_{+}/2)} +
I_{n-1}'^{+} \right)
+{(n+1) \over (n-1)}\;I_{n-1}^{+}
\left({I_{n+1}^{+} \over (M \tau_{+}/2)} +
I_{n+1}'^{+} \right)=0.
\eqno (A2)
$$
All terms in
(A2) have the same leading behaviour determined by exponential
functions. It is also known that non-exponential terms,
after taking logarithms and after expanding
these in inverse powers of $n$, are proportional
to inverse powers of $M$ and hence vanish in the limit $M
\rightarrow \infty$ [4]. Then equation (A.2) has the form:
$
{\rm exp. \; terms} \times
\left(1+{(n+1)\over (n-1)}\right)=0.
$
Moreover, it can be shown that also exponential terms do not
contribute to $I_{\rm pole}(\infty)$ [4].
Thus, one should only take the logarithm of
$\left(1+{(n+1)\over (n-1)}\right)$
and pick out the terms which,
multiplied by ${n^{2}\over 2}$,
yield terms proportional to $1/n$. This leads to the
result (3.6) for $I_{\rm pole}(\infty)$.

To calculate $I_{\rm pole}(0)$ it is more convenient to use again the
expression (A1) for the eigenvalue condition.
Taking the limit $M \rightarrow 0$, one finds the limiting
form of (A1) as
$$
{1 \over \Gamma(n+2)} \left({1 \over \Gamma(n-1)}
+{2 \over \Gamma(n)}\right)
+ {(n+1) \over (n-1)} {1 \over \Gamma(n)}
\left({1 \over \Gamma(n+1)} + {2 \over \Gamma(n+2)}\right)
=0.
\eqno (A3)
$$
Thus, inserting the asymptotic Stirling formula for
$\Gamma$-functions in the logarithm of the left-hand side
of equation (A3) and
expanding it in inverse powers of $n$, one finds the result (3.7) for
$I_{\rm pole}(0)$. For this purpose, we also use the expansion
of $\log(1+\omega)$ as $\omega \rightarrow 0$, i.e.
$
\log(1+\omega)=\sum_{k=1}^{\infty}(-1)^{k+1}{\omega^{k}\over k}
$ [1].

When one evaluates $I_{\rm pole}(\infty)$ and $I_{\rm pole}(0)$
in the Esposito gauge, it is also necessary to use the asymptotic
expansion of $\nu$ as $n \rightarrow \infty$ (cf (5.10) of [2])
$$
\nu \sim n \left(1-{3\over 8}{1\over n^{2}}
-{9\over 128}{1\over n^{4}}+{\rm O}(n^{-6})\right).
\eqno (A4)
$$
The contribution of gauge modes to $I_{\rm pole}(0)$ in
(2.17) is then obtained by taking the coefficient of
${1\over n}$ in the asymptotic expansion as
$n \rightarrow \infty$ of
$
{n^{2}\over 2} \log \left[{{(\nu-1/2)/(\nu+1/2)}\over
\Gamma(\nu+1/2)}\right]^{2}.
$
This yields the contribution
$$
I_{A}=-{9\over 128}-{1\over 32}+{1\over 360}
-{3\over 8}-{1\over 12}-{1\over 48}+{1\over 16}.
\eqno (A5)
$$
Moreover, the contribution of ghost modes to $I_{\rm pole}(0)$
in (2.17) is obtained by taking the coefficient of
${1\over n}$ in the asymptotic expansion as
$n \rightarrow \infty$ of
$
{n^{2}\over 2}\log
\left({1\over \Gamma(\nu+1)}\right)
$
and then multiplying by -2, since the ghost field is
fermionic and complex. Hence one finds the contribution
$$
I_{B}={9\over 128}+{1\over 32}-{1\over 360}.
\eqno (A6)
$$
Thus, the full $I_{\rm pole}(0)$ in (2.17) is given by
$$
I_{\rm pole}(0)=I_{A}+I_{B}
=-{5\over 12}.
\eqno (A7)
$$

When one evaluates $I_{\rm pole}(\infty)$ and
$I_{\rm pole}(0)$ for coupled gauge modes,
one finds that only $K$ functions at
$\tau=\tau_{-}$ and $I$ functions at $\tau=\tau_{+}$
contribute. The corresponding $I_{\rm pole}(0)$ values
are obtained as the coefficient of ${1\over n}$
when $n \rightarrow \infty$ and $M \rightarrow 0$
in the asymptotic expansions of the terms
$
{n^{2}\over 2}\log {(\nu-1/2)\over (\nu+1/2)},
{n^{2}\over 2}\log{(n+1)\over (n-1)},
{n^{2}\over 2}\log {(\nu-1/2)\over (\nu+1/2)}
$
respectively.
\vskip 0.3cm
\leftline {\bf References}
\vskip 0.3cm
\item {[1]}
Esposito G 1994 {\it Quantum Gravity, Quantum Cosmology and
Lorentzian Geometries} Lecture Notes in Physics, New
Series m: Monographs vol m12 second corrected and
enlarged edn (Berlin: Springer)
\item {[2]}
Esposito G 1994 {\it Class. Quantum Grav.} {\bf 11} 905
\item {[3]}
Louko J 1988 {\it Phys. Rev.} D {\bf 38} 478
\item {[4]}
Barvinsky A O, Kamenshchik A Yu and Karmazin I P 1992
{\it Ann. Phys., N.Y.} {\bf 219} 201
\item {[5]}
Hawking S W 1984 {\it Nucl. Phys.} B {\bf 239} 257
\item {[6]}
DeWitt B S 1965 {\it Dynamical Theory of Groups and Fields}
(New York: Gordon and Breach)
\item {[7]}
Moss I G and Poletti S J 1994
{\it Phys. Lett.} {\bf 333B} 326
\item {[8]}
Griffin P A and Kosower D A 1989 {\it Phys. Lett.}
{\bf 233B} 295
\item {[9]}
Kamenshchik A Yu and Mishakov I V 1994
{\it Phys. Rev.} D {\bf 49} 816
\item {[10]}
Abramowitz M and Stegun I A 1964
{\it Handbook of Mathematical Functions with Formulas, Graphs and
Mathematical Tables} (New York: Dover)
\item {[11]}
Vassilevich D V 1994 {\it Vector Fields on a Disk with Mixed
Boundary Conditions} (St Petersburg preprint SPbU-IP-94-6)

\bye